**Effect of annealing on the structural, topographical and optical properties of sol-gel derived ZnO and AZO thin films**

Joydip Sengupta[a*], R. K. Sahoo[b] and C. D. Mukherjee[c]

[a]Department of Physics, Sikkim Manipal Institute of Technology, Sikkim - 737136, India.

[b]Materials Science Centre, Indian Institute of Technology, Kharagpur 721302, India.

[c]ECMP Division, Saha Institute of Nuclear Physics,1/AF Bidhannagar, Calcutta 700064, India.

A comparative study of the physical properties of undoped Zinc Oxide (ZnO) and Al doped Zinc Oxide (AZO) thin films were performed as a function of annealing temperature. The structural properties were analyzed using X-ray diffraction and the recorded patterns indicated that the crystallinity of the films always enhanced with increasing annealing temperature while it degrades with Al doping. The topographical modification of the films due to heat treatment was examined by atomic force microscopy which revealed that annealing roughened the surface of all the films; however the AZO films always exhibited smoother morphology than ZnO. Study of optical properties by UV-Visible spectrophotometer demonstrated that the transmittance was gradually diminished with rise in annealing temperature. In addition, a notable increase in the optical bandgap was also observed for the AZO films.

**1. Introduction**

Transparent conductive oxides (TCOs) are extensively used in optoelectronic devices e.g. solar cells, flat panel displays etc. [1]. Though Indium Tin Oxide (ITO) is generally employed as TCO till date, a replacement is now required as Indium is toxic besides being expensive and scarce. Recently, ZnO is emerging as an alternative potential candidate to ITO owing to its cheap abundant raw material, direct band gap, large exciton binding energy, high transmittance in the visible region and non-toxic nature. Nevertheless, because of the high temperature instability of pure ZnO, presently doped ZnO is preferred for the potential applications [2]. Based on the previous literature, Al seems to be a successful and promising doping element for ZnO. Al can be substitutionally incorporated at the Zn lattice sites in the ZnO structure to fabricate AZO film with high temperature stability [3] and good immunity against hydrogen plasma reduction [4]. Additionally, AZO does not degrade the active solar cell materials because of the inter-diffusion of constituents as it occurs for ITO films [5]. However, optoelectronic devices require thermal treatments during fabrication and therefore TCOs must maintain its properties throughout such high temperature process. So the study of the effect of thermal treatment on the structural and optical properties of TCOs demands serious attention for optoelectronic applications of these films.

---

[*] Corresponding author: E-mail: joydipdhruba@gmail.com   FAX: +91-33-2337-4637

**2. Experimental procedure**

Zinc acetate dehydrate ($Zn(CH_3COO)_2 \cdot 2H_2O$), 2-Methoxyethanol and Monoethanolamine (MEA) were used as starting material, solvent and stabilizer, respectively. First, zinc acetate dihydrate was dissolved in 2-Methoxyethanol and then MEA was slowly added under magnetic stirring to prepare a solution of 0.75M. Aluminum doping was performed by adding aluminium nitrate nonahydrate ($Al(NO_3)_3 \cdot 9H_2O$) to the precursor solution in two different Al/Zn ratios, i.e. 0.8 and 2.0 at.%. The resulting mixture was stirred for 1 h at 65 °C, and then 3 h at room temperature to yield a clear and homogeneous solution. Then the solution was aged for 48 h at room temperature, which served as the coating solution. Thin films of ZnO and AZO were prepared by spin coating of the respective coating solutions onto pre-cleaned quartz substrates at rotation speed of 3000 rpm for 30 s in ambient condition. Afterwards the films were dried at 300 °C for 10 min in air to evaporate the solvent and organic residues. Finally these films were inserted into a furnace and annealed in air at 400, 550 and 700 °C for 1 h. An Atomic force microscope (AFM) (Nanonics Multiview 4000$^{TM}$) in intermittent contact mode, Philips X-ray diffractometer (XRD) (PW1729) with Cu source and UV-Visible (UV-Vis) spectrophotometer (PerkinElmer, Lambda 35) were used to characterize the samples.

## 3. Results and discussion

The XRD patterns of annealed ZnO and AZO thin films (Fig. 1) depicted that all films were polycrystalline in nature and of hexagonal wurtzite structure. Furthermore, for all the AZO films, neither $Al_2O_3$ nor Al phase was detected from the X-ray pattern. The diffraction pattern of ZnO film indicated a preferred orientation along (002) direction, while no such preferred growth orientation was observed for the AZO films. **The change of preferential growth direction is closely related to the alteration of surface free energy of the planes due to doping. The sufficient substitution of Zn ions by Al ions might modify the closely-packed surface, and made the surface energy value of the (002) plane higher or nearly equal to that of the other planes (e.g. (100) and (110)); being the reason for the change in the growth orientation [6].** The average grain size ($D$) of the films was also calculated using the full width at half maximum (FWHM) of (002) peak from the Scherrer's equation

$$D = \frac{K\lambda}{\beta \, Cos\theta} \qquad \ldots(1)$$

where $K = 0.9$ is the shape factor, $\lambda$ is the wavelength of incident X-ray, $\beta$ is the FWHM measured in radians and $\theta$ is the Bragg angle of diffraction peak. It was observed (Table 1) that, as the annealing temperature is raised; the FWHM value exhibited a tendency to decrease for ZnO as well as AZO films. The trend of FWHM values implied that the crystallinity of the films was improved with annealing [7]. Moreover, Al doping deteriorates the crystallinity of the films which may be caused by the stress formation as a result of the ion size difference between Al and Zn ($r_{Al}$ = 0.054 nm and $r_{Zn}$ = 0.074 nm).

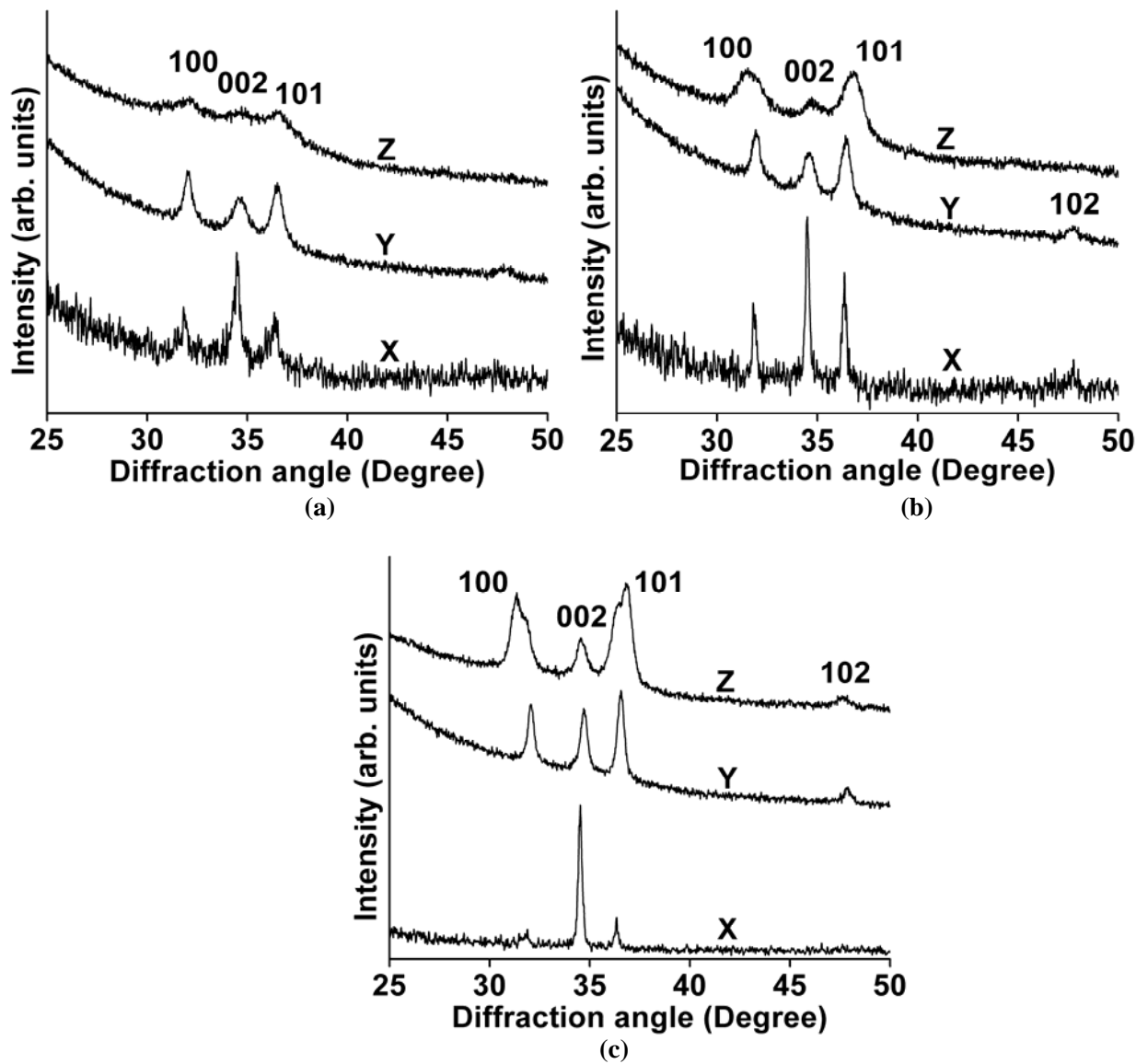

**Fig. 1. X-ray diffraction spectra of ZnO and AZO thin films deposited on quartz substrate after annealing at: (a) 400 °C, (b) 550 °C and (c) 700 °C (where X symbolize ZnO, Y symbolize 0.8 at.% AZO and Z symbolize 2.0 at.%. AZO)**

It was also observed (Table 1) that average grain size was increased with increasing annealing temperature for ZnO and AZO films. This could be explained by considering the thermal annealing induced coalescence of small grains by grain boundary diffusion which caused major grain growth [7]. **The grain growth mechanism includes the transfer of atoms at grain boundaries from one grain to another and the final grain size depends upon the specific** **annealing conditions. Grain size is a significant structural characteristic of thin films as carrier motilities are influenced by polycrystalline grain sizes [8].**

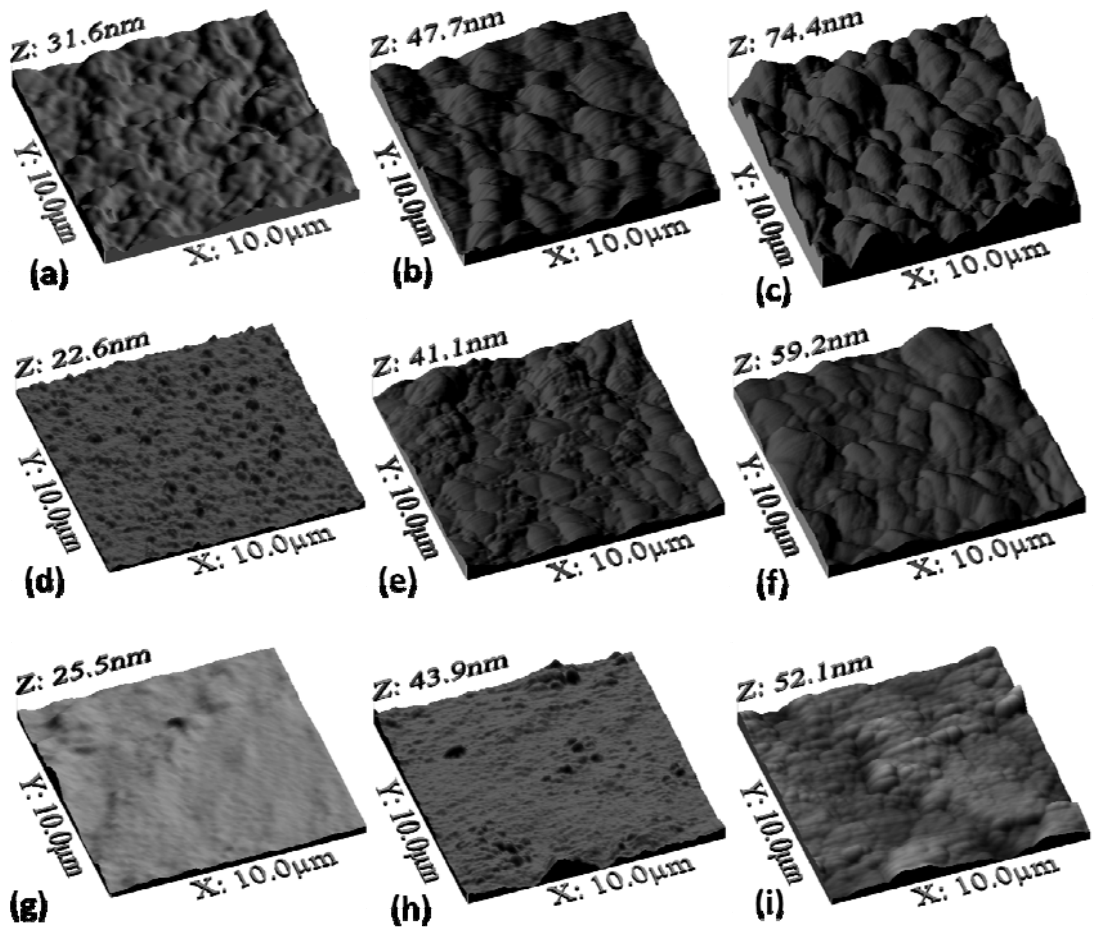

**Fig. 2.** Three dimensional AFM images of ZnO and AZO thin films after annealing at different temperatures in air for 1 h (a) 400 °C, (b) 550 °C, (c) 700 °C for ZnO film: (d) 400 °C, (e) 550 °C, (f) 700 °C for 0.8 at.% AZO film: (g) 400 °C, (h) 550 °C, (i) 700 °C for 2.0 at.% AZO film

AFM studies of ZnO and AZO thin films (Fig. 2) exhibits that all the films had tightly packed grains with good homogeneity and no cracks were observed. The root mean square (RMS) roughness of the annealed ZnO and AZO thin films was listed in Table 1, which revealed that RMS roughness of the AZO thin films is by far smaller than that of ZnO under same condition.

The decrease in surface roughness with Al doping may be caused by Al segregation at the non-crystalline region in the boundary [9]. Moreover, RMS roughness value of all films was increased with the increasing annealing temperature. This could be explained in terms of major grain growth which yields an increase in the surface roughness.

Spectrophotometric measurements of annealed ZnO and AZO thin films (Fig. 3) indicated that the transmittance had decreased with the increase in annealing temperature. It was previously reported that surface roughness strongly affects the transparency of ZnO-based thin films [10]. So it could be inferred that the major reason for the decrease in transmittance with higher annealing temperature may be due to the rough surface scattered and reflected light, as surface roughness increases upon annealing.

**Table 1.** The data evaluated form the XRD, AFM and UV-Vis measurements of sol-gel derived ZnO and AZO thin films after annealing at different temperatures

| Sample | Annealing Temperature (°C) | FWHM of (002) plane (degree) | Average grain size (nm) | RMS roughness (nm) | Optical Bandgap (eV) |
|---|---|---|---|---|---|
| ZnO | 400 | 0.3521 | 23 | 4.5 | 3.20 |
| | 550 | 0.2281 | 36 | 6.7 | 3.17 |
| | 700 | 0.2089 | 39 | 10 | 3.15 |
| 0.8 at.% AZO | 400 | 0.7299 | 11 | 2.5 | 3.25 |
| | 550 | 0.5841 | 14 | 5.3 | 3.24 |
| | 700 | 0.4107 | 20 | 8.1 | 3.23 |
| 2.0 at.%. AZO | 400 | 1.1460 | 7 | 2 | 3.38 |
| | 550 | 0.9094 | 9 | 4.4 | 3.35 |
| | 700 | 0.5315 | 15 | 6.5 | 3.32 |

The optical band gap of ZnO and AZO films could be estimated by employing the Tauc model:

$$(\alpha h\nu) = A(h\nu - E_g)^{1/2} \quad \ldots (2)$$

where $\alpha$ is the absorption coefficient, $h\nu$ is the photon energy, $A$ is a constant and $E_g$ is the optical bandgap. The optical bandgap of the thin films annealed at different temperatures was determined by extrapolation of the straight section to the energy axis of the plot of $(\alpha h\nu)^2$ versus photon energy (insets of Fig. 3). Table 1 lists the extrapolated bandgap values of all the samples

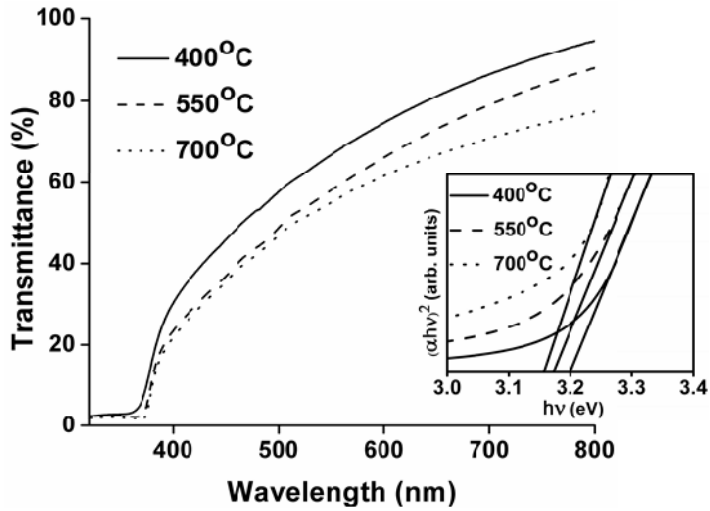

(a)

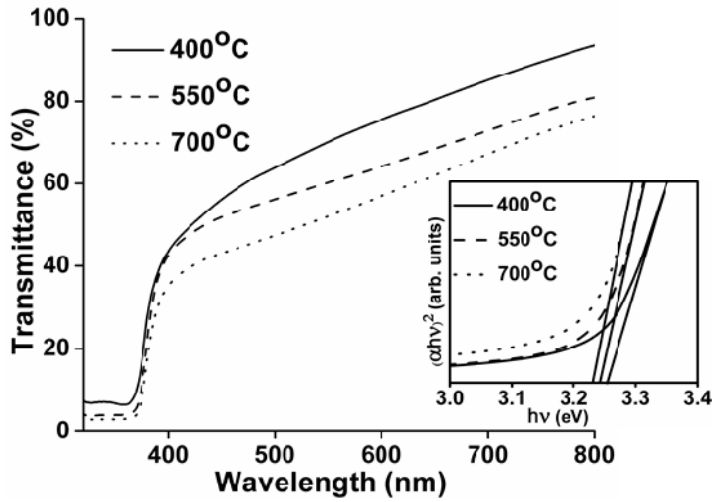

(b)

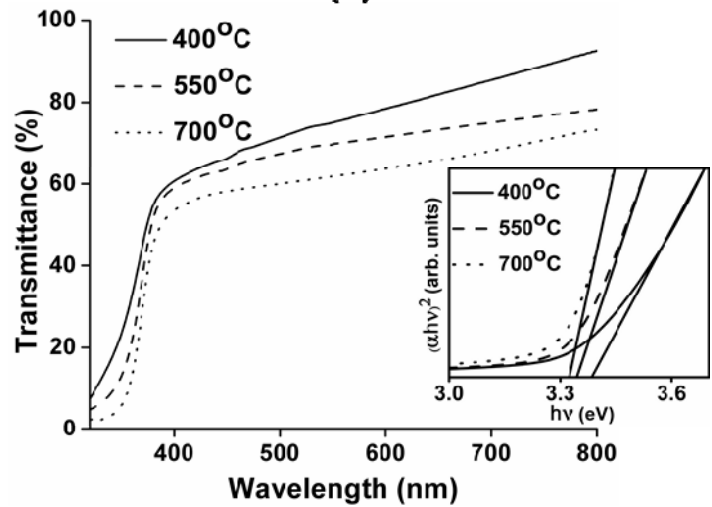

(c)

which showed decrease in bandgap with increasing annealing temperature. The shifts of the optical band gap might be attributed to the decreased defect of the thin films with the annealing temperature. Moreover, AZO thin films always exhibits higher optical band gap than ZnO and it increases with Al doping concentration (Table 1). We attribute the higher band gap of AZO thin films to the Burstein–Moss effect caused by an increase in free electron concentration due to the Al doping.

**Fig. 3.** Optical transmittance spectra of sol-gel derived ZnO and AZO thin films after annealing at different temperatures. (Inset) Tauc's plot of annealed films on quartz substrate. (a) ZnO, (b) 0.8 at.% AZO, (c) 2.0 at.% AZO

## 4. Conclusions

We have investigated the structural, topographical and optical properties of sol-gel derived ZnO and AZO thin films deposited on quartz substrate with respect to annealing. The study revealed that the crystallinity along with average grain size of all the films was increased with annealing temperature. However, AZO films exhibited poor crystallinty and also lower surface roughness in comparison to ZnO. The investigation also showed that RMS roughness of both ZnO and AZO films increases upon annealing which in effect reduces the transmittance. Furthermore, appreciable amount of increase in optical band gap was observed via Al doping depicting the Burstein–Moss effect.